\documentclass[11pt,preprint]{aastex}

\usepackage{lscape}

\begin{document}

\shorttitle{NLTE effects on AGB stars in 47 Tucanae}
\shortauthors{Lapenna et al.}

\title{NLTE effects on the iron abundance of AGB stars in 47
  Tucanae\footnotemark[1]} \footnotetext[1]{Based on UVES, FLAMES and
  FEROS observations collected under Programs 073.D-0211, 074.D-0016
  074.D-0571 and 090.D-0153.}

\author{
E. Lapenna\altaffilmark{2},
A. Mucciarelli\altaffilmark{2},
B. Lanzoni\altaffilmark{2},
F. R. Ferraro\altaffilmark{2},
E. Dalessandro\altaffilmark{2},
L. Origlia\altaffilmark{3},
D. Massari\altaffilmark{2}
}
\affil{\altaffilmark{2} Dipartimento di Fisica e Astronomia, Universit\`a degli
  Studi di Bologna, Viale Berti Pichat 6/2, I--40127 Bologna, Italy}
\affil{\altaffilmark{3} INAF- Osservatorio Astronomico di Bologna, Via
  Ranzani, 1, 40127 Bologna, Italy}

\date{10 October 2014}

\begin{abstract}
We present the iron abundance of 24 asymptotic giant branch (AGB)
stars members of the globular cluster 47 Tucanae, obtained with
high-resolution spectra collected with the FEROS spectrograph at the
MPG/ESO-2.2m Telescope.  We find that the iron abundances derived from neutral
lines (with mean value [Fe I/H]$ = -0.94 \pm 0.01$, $\sigma= 0.08$ dex)
are systematically lower than those derived from single ionized lines
([Fe II/H]$ = -0.83 \pm 0.01$, $\sigma= 0.05$ dex). Only the latter are
in agreement with those obtained for a sample of red giant branch
(RGB) cluster stars, for which Fe I and Fe II lines provide the same
iron abundance.  This finding suggests that Non Local Thermodynamical
Equilibrium (NLTE) effects driven by overionization mechanisms are
present in the atmosphere of AGB stars and significantly affect Fe I
lines, while leaving Fe II features unaltered. On the other hand, the
very good ionization equilibrium found for RGB stars indicates that these NLTE
effects may depend on the evolutionary stage.  We discuss the impact
of this finding both on the chemical analysis of AGB stars, and on the
search for evolved blue stragglers.
\end{abstract}


\section{Introduction}
\label{intro}
The asymptotic giant branch (AGB) evolutionary phase is the final
stage of thermonuclear burning for low/intermediate mass stars (M $<$ 8~$M_\odot$).
AGB stars are known to dominate the integrated light of
intermediate age stellar systems 
\citep[see, e.g.,][]{renzini86,fe95,hoefner98,vanloon99,cioni03,maraston05}
and to play a crucial role in the chemical enrichment processes of
cosmic structures. In fact, from one side they are the main sites
where elements heavier than the Fe-peak (like Ba and La) are formed,
through slow neutron captures \citep{busso99}, and promptly ejected.
On the other side,
their interiors can reach temperatures high enough to activate the
proton-capture reaction chains, like the NeNa and the MgAl ones.
Intermediate-mass AGB stars are also considered to be responsible for
the chemical anomalies (involving C, N, O, Na, Mg and Al) detected in
all properly observed Galactic globular clusters (GCs; see
\citealp{gratton12}) and in old Magellanic Cloud globulars
\citep{mucciarelli09}.

Due to their short lifetimes (of the order of $\sim$10 Myr for a
low-mass star with 0.8~$M_\odot$), the number of AGB stars observable
in an old GC is small, about a factor of 3-4 smaller than the number
of red giant branch (RGB) stars of comparable
luminosity. Additionally, high-quality photometry and large color
baselines are needed to properly separate genuine AGB from RGB stars.
Hence, despite their luminosity, these objects are still not well
investigated from the chemical point of view. Accurate determinations
of their atmospheric parameters and metallicity based on
high-resolution spectroscopy are available for a few GCs only, namely
M~4 \citep{ivans99}, M~5 \citep{ivans01, koch10} and 47 Tucanae
\citep[hereafter 47Tuc,][]{wylie06,koch08,worley09}, and they are all based on small
samples (no more than 6 AGB stars in each of the quoted
clusters).  Particular care has been addressed to the different
distribution of CN \citep{mallia78,norris81,briley93} and Na
\citep{campbell13} in AGB and RGB stars, in light of the
self-enrichment processes predicted to occur in the early stage of GC
life \citep[see e.g.][]{dercole08}.

New interest for AGB stars has been raised in a different research
field from the observational results of \citet{beccari06}, who found a
significant excess ($\sim 30\%$) and an anomalous segregation of such
objects in the central regions of 47 Tucanae
(see also \citealt{dalessandro09} for the case of M~2).
This finding suggests that the genuine AGB population of this GC may be
``contaminated'' by a peculiar class of heavy intruders, namely the
descendants of blue straggler stars\footnote{Blue stragglers are a
peculiar class of stars located along an extrapolation of the main
sequence, brighter and bluer than the turnoff point, where objects
of 1.2-1.5~$M_\odot$ spend their core hydrogen burning phase. They
are thought to be more massive than normal cluster stars
\citep[e.g.][]{shara97,gilliland98,lan07_1904,fiorentino14} and to be
generated by stellar interactions and/or mass transfer in binary
systems \citep{mccrea64,hills76}. Hence they are powerful tracers of
dynamical evolution \citep[e.g.][]{fe09_m30,fe12}.}  experiencing
their giant evolutionary phases (hereafter, evolved blue straggler
stars, e-BSSs). While confirming such a hypothesis would provide
precious constraints to the evolutionary models of these exotica, no
firm identification of even a single e-BSS has been obtained to
date. Indeed, a detailed investigation of AGB stars could finally fill
this gap.

In this paper we present the determination of the iron abundance for a
sample of 24 AGB stars in the metal-rich Galactic GC 47 Tuc.  The
observations and the spectroscopic analysis are described in Section
\ref{obs} and \ref{analysis}, respectively. Section \ref{results}
presents the measured iron abundances and a series of tests
demonstrating the correctness of the adopted procedure and
assumptions. A discussion of the results obtained and of their impact
on the traditional chemical analysis and the search for e-BSSs is
provided in Section \ref{discussion}. The conclusions of the work are
summarized in Section \ref{summary}.

\section{Observations}
\label{obs}
\begin{figure}[]
\includegraphics[trim=0cm 0cm 0cm 0cm,clip=true,scale=.60,angle=270]{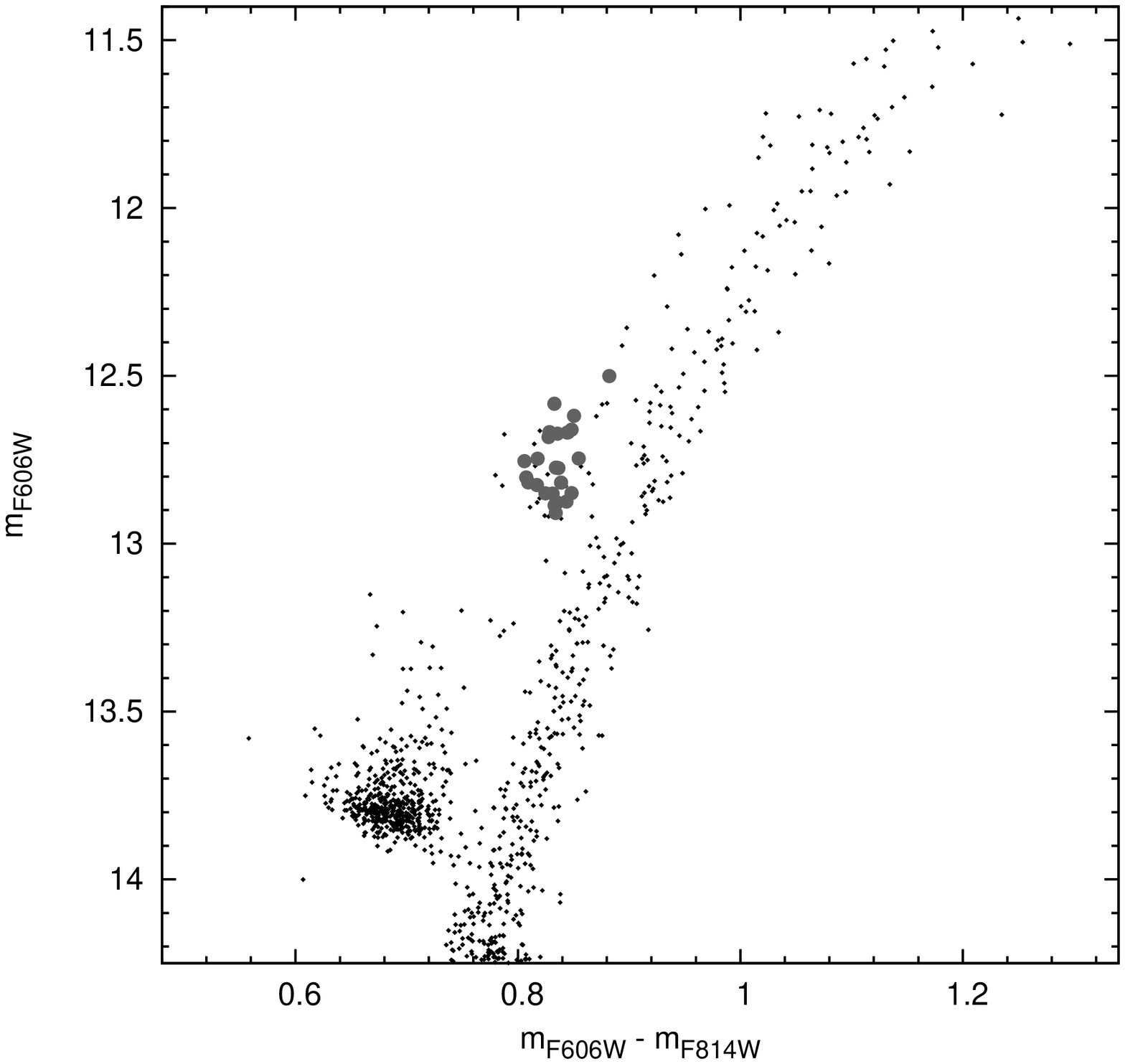}
\includegraphics[trim=0cm 0cm 0cm 0cm,clip=true,scale=.60,angle=270]{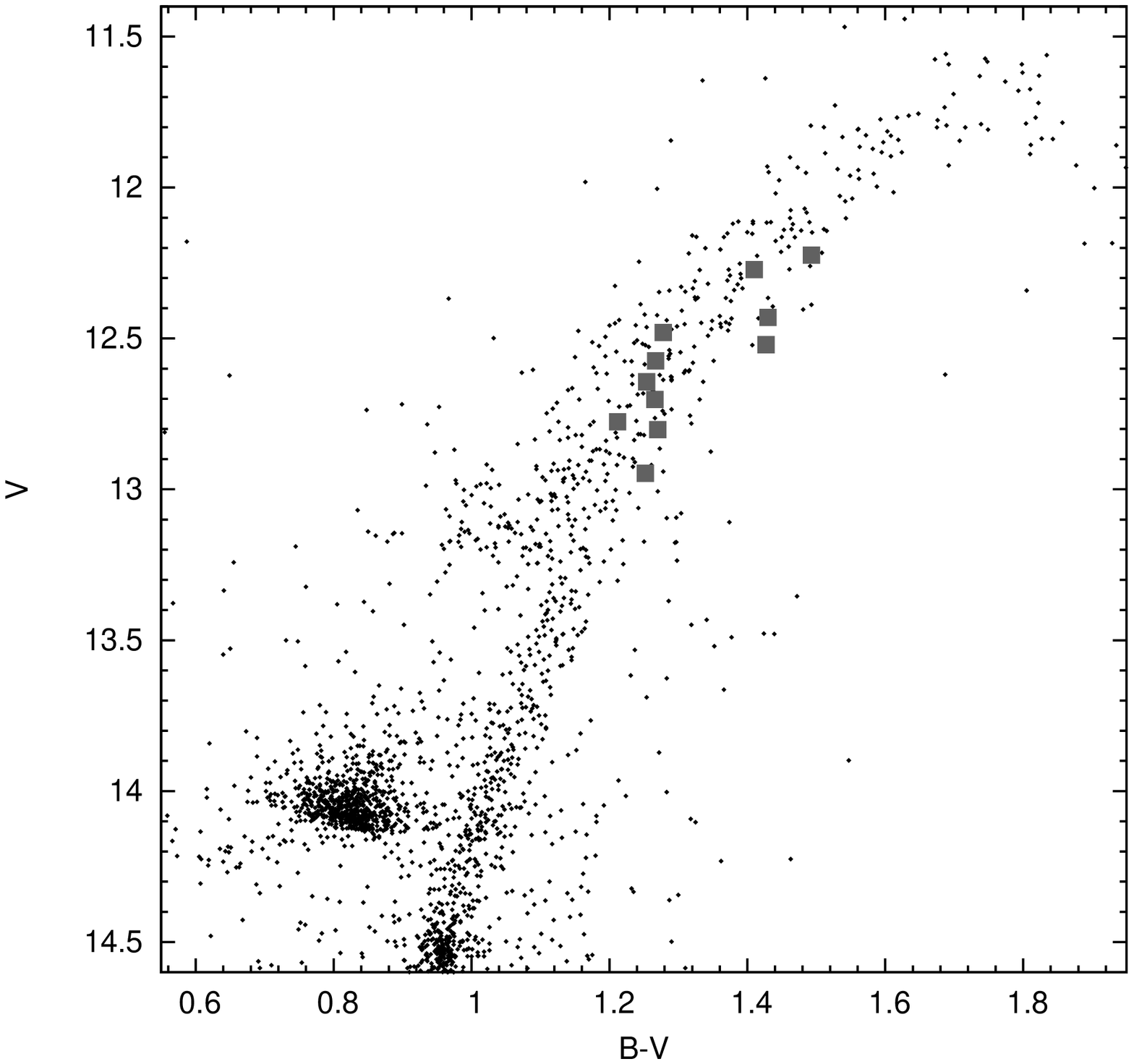}
\caption{Top panel: $(m_{\rm F606W},m_{\rm F606W}-m_{\rm F814W})$ color magnitude diagram of 47 Tuc from
\citet{beccari06}. The large solid circles mark the 24 targets studied in the present work.
Bottom panel: ($V,V-I$) color magnitude diagram of 47 Tuc obtained from WFI data by \citet{ferraro04}.
The large solid squares mark the 11 RGB stars used for comparison.}
\label{cmd}
\end{figure}

We acquired high resolution spectra of 24 AGB stars in 47 Tuc (Program
ID 090.D-0153, PI: Lanzoni) by using the Fibre-fed Extended Range
Optical Spectrograph \citep[FEROS;][]{kaufer99} mounted at the
MPG/ESO-2.2m telescope.  The spectra cover a wavelength range between
$\lambda \sim 3500~\mathring{\rm A}$ and $\lambda \sim 9200
~\mathring{\rm A}$, with a spectral resolution of $\sim 48000$. FEROS
allows to allocate simultaneously two fibers at a relative distance of
$2.9\arcmin$, one on the source and the other on the sky.  The targets
have been selected from the photometric catalog of \citet{beccari06},
within $\sim 100\arcsec$ from the cluster center. In the
color-magnitude diagram (CMD) of 47 Tuc they are located in the AGB
clump (corresponding to the beginning of this evolutionary phase;
\citealp{fe99}), at $m_{\rm F606W} \sim 12.8$ and
$(m_{\rm F606W}-m_{\rm F814W})\simeq 0.8$ (see top panel of Figure \ref{cmd}).
Only isolated stars have been selected, in order to avoid
contamination of the spectra from close objects of larger or comparable
luminosity.  The identification number, coordinates and magnitudes of
each target are listed in Table \ref{tab1}.

For each target a single exposure of $\sim$30-40 min has been
acquired, reaching signal-to-noise ratios S/N$\geq 70$ per pixel.
The data reduction was performed by using the ESO FEROS pipeline,
including bias subtraction, flat fielding, wavelength calibration
by using a Th-Ar-Ne reference lamp, spectrum extraction and final
merging and rebinning of the orders.  Since the background level of
the sky is negligible ($<1\%$) compared to the brightness of the
observed targets, we did not perform the sky subtraction from the
final spectra in order to preserve its maximum quality.  We accurately
checked that the lack of sky subtraction has no impact on the derived
abundances, by comparing the equivalent widths (EWs) measured 
for some spectra with and without the sky subtraction.

\section{Analysis}
\label{analysis}

\subsection{Radial velocities}
\label{vrad}
The radial velocity of each target has been obtained by means of
the code DAOSPEC \citep{stetson08}, measuring the positions of more than 300
metallic lines. The accuracy of the wavelength calibration has been
checked by measuring telluric absorptions and oxygen sky lines,
finding no significant zero-point offsets. Uncertainties have
been computed as the dispersion of the measured radial velocities
divided by the square root of the number of used lines, and they
turned out to be smaller than 0.04 km s$^{-1}$.
Heliocentric corrections obtained with the IRAF task RVCORRECT have
been adopted.  The heliocentric radial velocities for all the targets
are listed in Table \ref{tab1}. They range between $\sim -41.5$ and
$\sim +9.5$ km s$^{-1}$, with mean value of $-17.6 \pm 2.3$ km
s$^{-1}$ and dispersion $\sigma = 11.5$ km s$^{-1}$.  These values are
in good agreement with previous determinations of the systemic radial
velocity of 47 Tuc \citep[see e.g.][]{mayor83,meylan91,gebhardt95,carretta04,alvesbrito05,koch08,lane10}.
All the targets have been considered as members of the
cluster, according to their radial velocities and distance from the
cluster center.


\subsection{Chemical analysis}
\label{chem}
The chemical abundances have been derived by using the package
GALA\footnote{http://www.cosmic-lab.eu/gala/gala.php}
\citep{mucciarelli13} which matches the measured and the theoretical
equivalent widths \citep[see][for a detailed description of this
  method]{castelli05}.  The model atmospheres have been computed by
using the ATLAS9 code, under the assumption of plane-parallel
geometry, local thermodynamical equilibrium (LTE) and no overshooting
in the computation of the convective flux.  We adopted the last
release of the opacity distribution functions from \citet{castelli04},
assuming a global metallicity of [M/H] = $-1$ dex with [$\alpha$/Fe]$ = +0.4$ dex
for the model atmospheres.

The effective temperatures ($T_{\rm eff}$) and surface gravities
($\log g$) of the targets have been derived photometrically, by
projecting the position of each star in the CMD onto the isochrone
best fitting the main evolutionary sequences of 47 Tuc. The isochrone
has been extracted from the BaSTI database \citep{pietrinferni06}
assuming an age of 12 Gyr, metallicity Z $= 0.008$ and $\alpha$-enhanced
chemical mixture. We adopted a distance modulus $(m-M)_V = 13.32$ mag
and a color excess $E(B-V) = 0.04$ mag \citep{fe99}.  Microturbulent
velocities ($v_{\rm turb}$) have been derived by requiring that no
trends exist between Fe I abundances and the reduced EWs, defined as
$\log(EW/\lambda)$. The adopted atmospheric parameters are listed in
Table \ref{tab2}.

Only absorption lines that are predicted to be unblended at the FEROS
resolution have been included in our analysis. The line selection has
been performed through a careful inspection of synthetic spectra
calculated with the code SYNTHE \citep{sbordone05} assuming the
typical atmospheric parameters of our targets and the typical
metallicity of 47 Tuc.
We considered only transitions with accurate theoretical/laboratory
atomic data taken from the last version of the Kurucz/Castelli
compilation.\footnote{http://wwwuser.oat.ts.astro.it/castelli/linelists.html}
The EWs have been obtained with DAOSPEC \citep{stetson08}, iteratively
launched by means of the package
4DAO\footnote{http://www.cosmic-lab.eu/4dao/4dao.php}\citep{mucciarelli13b}
that allows an analysis cascade of a large sample of stellar spectra
and a visual inspection of the Gaussian fit obtained for all the
investigated lines.  Due to the extreme crowding of spectral lines in
the region between $\lambda \sim 3800~\mathring{\rm A}$ and $\lambda
\sim 4500~\mathring{\rm A}$, and to the presence of several absorption
telluric line bands beyond $\sim 6800~\mathring{\rm A}$, we restricted
the analysis to the spectral range between $\sim 4500~\mathring{\rm
  A}$ and $\sim 6800~\mathring{\rm A}$.  In order to avoid too weak or
saturated features, we considered only lines with reduced EWs between
$-5.6$ and $-4.7$ (these correspond to EW = 11 m$\mathring{\rm A}$ and
90 m$\mathring{\rm A}$ at $\lambda\sim 4500~\mathring{\rm A}$, and EW
= 17 m$\rm\mathring{A}$ and 135 m$\rm\mathring{A}$ at $\lambda \sim
6800~\mathring{\rm A}$, respectively).  Moreover, we discarded from
the analysis also the lines with EW uncertainties larger than $20\%$,
where the uncertainty of each individual line is provided by DAOSPEC
on the basis of the fit residuals. With these limitations, the iron
abundance has been derived, on average, from $\sim 150$ Fe I lines and
$\sim 13$ Fe II lines.
In the computation of the final iron abundances we adopted
as reference solar value A(Fe)$_{\odot}$ = 7.50 dex \citep{grevesse98}.

Uncertainties on the derived abundances have been computed for each
target by adding in quadrature the two main error sources: {\sl (a)}
those arising from the EW measurements, which have been estimated as
the line-to-line abundance scatter divided by the square root of the
number of lines used, and {\sl (b)} the uncertainties arising from the
atmospheric parameters, computed varying by the corresponding
uncertainty only one parameter at a time, while keeping the others
fixed. The abundance variations thus obtained have been added in
quadrature.
Term (a) is of the order of less than 0.01 dex for Fe~I and 0.03 dex for Fe~II.
Since the atmospheric parameters have been estimated from photometry,
by projecting the position of each target in the CMD onto the isochrone,
we estimated term (b) from the photometric uncertainties.
By assuming a conservative uncertainty of 0.1 mag for the magnitudes of our targets
we obtained an uncertainty of about $\pm$50 K and $\pm$0.05 dex
on the final T$_{eff}$ and log~$g$, respectively.
The total uncertainties in [Fe I/H] are of the order of
0.04-0.05 dex, while in [Fe II/H] are of about 0.08-0.10 dex (due to
the higher sensitivity of Fe II lines to $T_{\rm eff}$ and $\log g$).

\section{Iron abundance}
\label{results}
\begin{figure}[]
\includegraphics[trim=0cm 0cm 0cm 0cm,clip=true,scale=.70,angle=0]{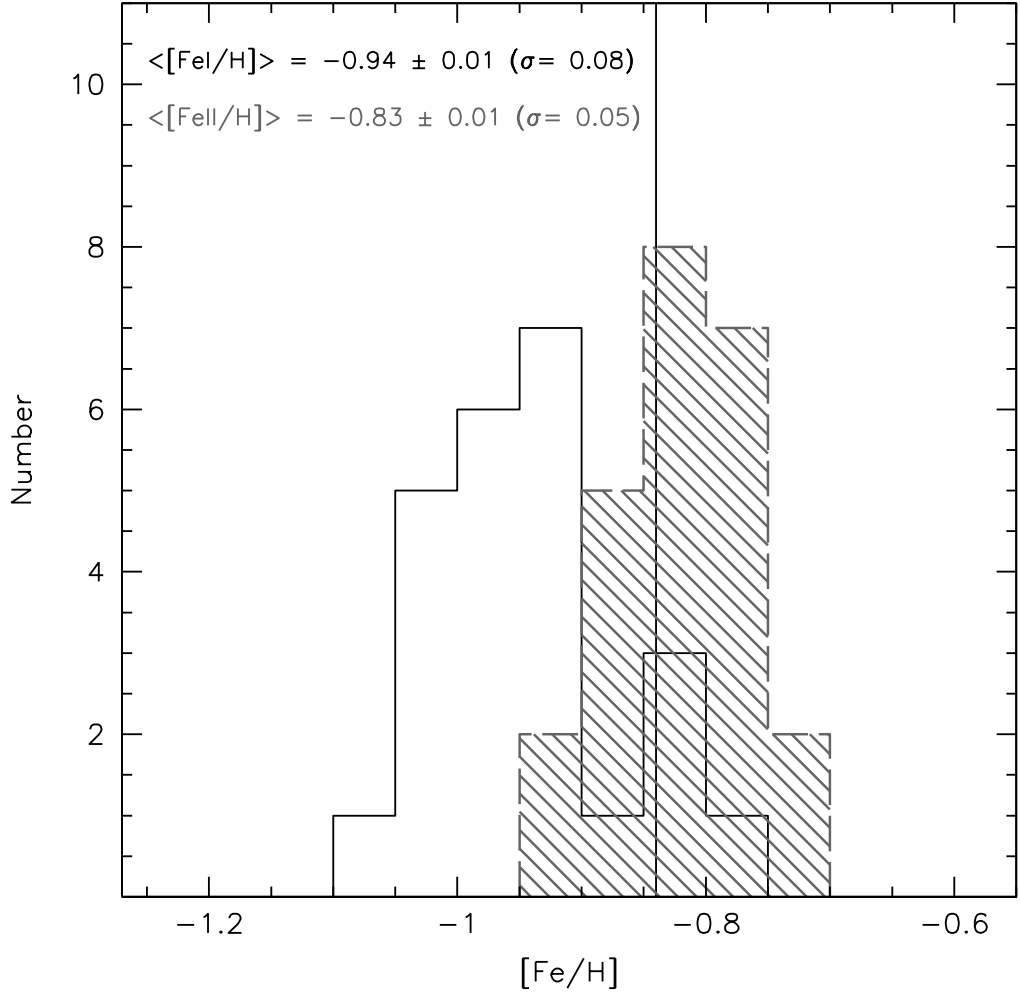}
\caption{Distribution of the iron abundance ratios measured for the 24
  AGB stars in our sample, from Fe I lines (empty histogram) and from
  Fe II lines (shaded histogram).  The vertical line indicates the
  average iron abundance derived from 11 RGB stars.}
\label{iron}
\end{figure}

\begin{figure}[]
\includegraphics[trim=0cm 3.0cm 0cm 3.0cm,clip=true,scale=.70,angle=270]{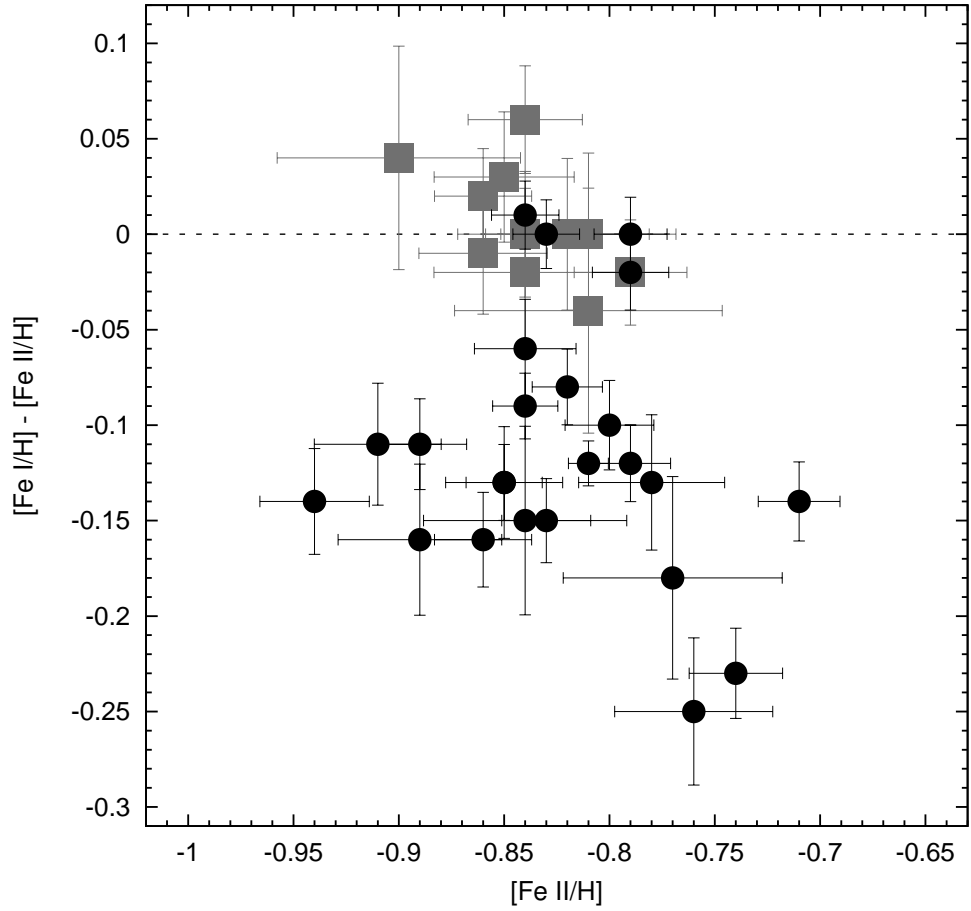}
\caption{Difference between [Fe I/H] and [Fe II/H] as a function of
  [Fe II/H] for the 24 AGB stars (black circles) and the 11 RGB stars
  (gray squares) in our 47 Tuc samples. }
\label{fediff}
\end{figure}

The [Fe I/H] and [Fe II/H] abundance ratios measured for each target
are listed in Table \ref{tab2}, together with the total uncertainties
and the number of lines used.  Their distributions are shown in Figure
\ref{iron}.  A systematic difference between [Fe I/H] and [Fe II/H] is
evident, with the abundances derived from Fe I lines being, on
average, $0.1$ dex smaller than those obtained from Fe II: the mean
values of the distributions are [Fe I/H] $= -0.94 \pm 0.01$ ($\sigma =
0.08$ dex) and [Fe II/H] $= -0.83 \pm 0.01$ ($\sigma = 0.05$ dex).
These values are clearly incompatible each other. Moreover,
only [FeII/H] is in agreement with the metallicity quoted in the literature
and based on sub-giant or RGB stars
\citep{carretta04,alvesbrito05,koch08,carretta09}, while the iron
abundance obtained from Fe I lines is significantly smaller. The
distribution of the difference [Fe I/H]$-$[Fe II/H] is shown as a
function of [Fe II/H] in Figure \ref{fediff} (black circles). It is quite broad,
ranging from $-0.25$ to $+0.01$ dex.

\subsection{Sanity checks}
\label{check}
The difference in the derived [Fe I/H] and [Fe II/H] abundances cannot be
easily explained (especially if considering the high quality of the acquired spectra and the very
large number of used iron lines) and it clearly needs to be understood.
In order to test the correctness of our analysis and to exclude
possible bias or systematic effects, we therefore performed a number
of sanity checks.

\subsubsection{Checks on the chemical analysis procedure}
To test the reliability of our chemical analysis, we
studied a sample of RGB stars in 47 Tuc, the standard star 104 Tau,
Arcturus and the Sun by following the same procedure adopted for the
AGB targets discussed above (i.e., by using the same linelist, model
atmospheres, and method to infer the atmospheric parameters).

\emph{RGB stars in 47 Tuc --} We measured the iron abundance for a
sample of 11 RGB stars in 47 Tuc, for which high-resolution
(R$\sim$45000) FLAMES-UVES spectra are available in the ESO Archive
(Program ID: 073.D-0211).
The location of these stars in the ($V$,$V-I$) CMD from \citet{ferraro04}
is shown in Figure \ref{cmd} (bottom panel).
Their atmospheric parameters and [Fe~I/H] and [Fe~II/H] abundance ratios
are listed in Table \ref{tab3}.
The distribution of the [Fe~I/H]-[Fe~II/H] differences is shown
in Figure \ref{results} (large gray squares).
We found an average [Fe I/H]$_{\rm RGB} = -0.83 \pm
0.01$ dex ($\sigma= 0.02$ dex) and [Fe II/H]$_{\rm RGB} = -0.84 \pm
0.01$ dex ($\sigma= 0.03$ dex).
These values are fully consistent with previous determinations.
In fact, the careful comparison with the results of \citet{carretta09}, who
analyzed the same 11 RGB spectra, shows that the average differences
in the adopted parameters are: $\Delta T_{\rm eff} = 36 \pm 8$ K
($\sigma$ = 26 K), $\Delta \log g = 0.07 \pm 0.01$ ($\sigma$ = 0.04),
$\Delta v_{\rm turb} = 0.01 \pm 0.04$ km s$^{-1}$ ($\sigma$ = 0.14 km
s$^{-1}$).  By taking into account also the differences among the
adopted atomic data, model atmospheres and procedure to measure the
EWs, the derived abundances turn out to be in very good agreement
within the uncertainties, the mean difference between our values and
those of \citet{carretta09} being $\Delta$[Fe I/H]$ = -0.06 \pm 0.03$
dex and $\Delta$[Fe II/H]$ = -0.02 \pm 0.03$ dex.


\emph{104 Tau --} We applied the same procedure of data reduction and
spectroscopic analysis to the star 104 Tau (HD 32923), which was
observed during the first observing night as a radial velocity
standard star. The spectrum has been reduced with the same set of
calibrations (i.e., bias, flat-fields, Th-Ar-Ne lamp) used for the
main targets.  We obtained a radial velocity of $20.96 \pm 0.02$ km
s$^{-1}$, which is consistent with the value ($20.62 \pm 0.09$ km
s$^{-1}$) quoted by \citet{nidever02}.  We adopted the average values
of $T_{\rm eff}$ and $\log g$ provided by \citet{takeda05} and
\citet{ramirez09} in previous analyses of the star ($T_{\rm eff}$=
5695 K and $\log g= 4.05$), while $v_{\rm turb}$ has been constrained
spectroscopically.  We derived [Fe I/H] $= -0.17 \pm 0.01$ ($\sigma$ =
0.09 dex) and [Fe II/H] $= -0.20 \pm 0.02$ ($\sigma$ = 0.07 dex), well
matching, within a few hundredth of dex, the values quoted by
\citet{takeda05} and \citet{ramirez09}.


\emph{Arcturus and the Sun --} In order to test the robustness of the
used linelist (and in particular to check for possible systematic
offsets due to the adopted oscillator strengths of Fe I and Fe II
lines), we adopted the same procedure to measure the iron abundance of
Arcturus (HD 124897) and the Sun, both having well established
atmospheric parameters.  In the case of Arcturus we retrieved a FEROS
spectrum from the ESO archive (Program ID: 074.D-0016), adopting 
$T_{\rm eff}$ = 4300 K, $\log g$ = 1.5, $v_{\rm turb}$ = 1.5 km s$^{-1}$
and [M/H]$ = -0.5$ as derived by \cite{lecureur07}.
We obtained [Fe I/H]$ = -0.56 \pm 0.01$
dex and [Fe II/H]$ = -0.57 \pm 0.01$ dex, in very good agreement with
previous determinations \citep{fulbright06, lecureur07, ramirez11}.
We repeated the same test on a FLAMES-UVES ($R\sim 45000$) twilight
spectrum of the
Sun\footnote{http://www.eso.org/observing/dfo/quality/GIRAFFE/pipeline/solar.html},
adopting $T_{\rm eff}$ = 5777 K, $\log g$ = 4.44 and $v_{\rm turb}$ =
1.0 km s$^{-1}$ and finding absolute Fe abundances of 7.49 $\pm$ 0.01
dex and 7.50 $\pm$ 0.02 dex from neutral and single ionized iron lines,
respectively.

\subsubsection{Checks on the atmospheric parameters}

\emph{Effective temperature and gravity from spectroscopy --} We
checked whether atmospheric parameters derived spectroscopically could
help to reconcile the [Fe I/H] and [Fe II/H] abundance ratios.  The
adopted photometric estimates of $T_{\rm eff}$ well satisfy the
{\it excitation balance} (i.e.  there is no slope between abundances and
excitation potential). Hence very small (if any) adjustments, with a
negligible impact on the derived abundances, can be admitted.
Instead, the values adopted for the surface gravity can have a
significant impact on the difference between [Fe I/H] and [Fe II/H].
In fact, the abundances derived from Fe II lines are a factor of $\sim
5$ more sensitive to variations of $\log g$ with respect to those
obtained from Fe I lines: for instance, a variation of $-0.1$ dex in
$\log g$ leads to negligible variation ($\sim -0.01$ dex) in the Fe I
abundance, while [Fe II/H] decreases by 0.05 dex.  Hence, a lower
value of the surface gravity could, in principle, erase the difference
between [Fe I/H] and [Fe II/H].

We found that the derived spectroscopic gravities are on average lower
than the photometric ones by 0.25 dex, with a maximum difference of
$\sim$0.5 dex.  As an example, the photometric analysis of star \#100103
provides [Fe I/H]$ = -1.01 \pm 0.01$ ($\sigma$ = 0.11 dex) and [Fe
  II/H] $= -0.76 \pm 0.04$ ($\sigma$ = 0.13 dex), with $T_{\rm eff}$ =
4450 K, $\log g$ = 1.60 and $v_{\rm turb}$ = 1.15 km s$^{-1}$. When a
fully spectroscopic analysis is performed (thus optimizing all the
atmospheric parameters simultaneously), we obtain $T_{\rm eff}$ = 4475
K, $\log g$ = 1.10 and $v_{\rm turb}$ = 1.25 km s$^{-1}$, and the
derived abundances are [Fe I/H] $= -1.09 \pm 0.01$ and [Fe II/H] =
$-1.10 \pm 0.02$.  As expected, the spectroscopic values of $T_{\rm
eff}$ and $v_{\rm turb}$ are very similar to those adopted in the
photometric analysis.  A large difference is found for $\log g$, but
the final iron abundances are both too low with respect to the
literature and the 11 RGB star values to be considered acceptable.
Similar results are obtained for all the other AGB targets in our
sample.

Small surface gravity values as those found from the fully
spectroscopic analysis would require that stars reach the AGB phase
with an average mass of 0.4~$M_\odot$ (keeping $T_{\rm eff}$ and
luminosity fixed). This is lower than the value expected by
considering a main sequence turnoff mass of 0.9~$M_\odot$ and a $\sim
0.25~M_\odot$ mass loss during the RGB phase \citep{origlia07,origlia14}. 
Note that for some stars where [Fe I/H]$-$[Fe II/H]$\le
-0.20$ dex, the derived spectroscopic gravity would require a stellar
mass of $\sim 0.2~M_\odot$, which is even more unlikely for GC stars
in this evolutionary stage, also by taking into account the mass loss
rate uncertainties \citep{origlia14}.  Alternatively, these values of
$\log g$ can be obtained by assuming a significantly larger (by
about 0.5 mag) distance modulus.  However, this would be incompatible
with all the previous distance determinations for 47 Tuc \citep[see
  e.g.][]{fe99,McLaughlin,bergbuch09}.

\emph{Effective temperature and gravity from a different photometric
approach --} We repeated the analysis by adopting effective
temperatures estimated from the de-reddened color of each
target and the $(V-I)_0$-$T_{\rm eff}$ relation provided by
\cite{alonso99}, based on the Infrared Flux Method \citep[see][and
  references therein]{blackwell90}.  Because this color-$T_{\rm eff}$
relation is defined in the Johnson photometric system, we converted
the target magnitudes in that system, following the prescriptions of
\citet{sirianni05}.  Moreover, gravities have been computed from the
Stefan-Boltzmann relation, by using the derived values of $T_{\rm eff}$,
the luminosities obtained from the observed $V$-band
magnitudes, assuming a mass of 0.8~$M_\odot$ for all the
stars (according to the best-fit isochrone discussed above) and
adopting the bolometric corrections computed according to
\citet{buzzoni10}. The average difference between the $T_{\rm eff}$
values obtained from the isochrone and those derived from the
\cite{alonso99} relation is of 3 K ($\sigma$ = 50 K). For gravities we
obtained an average difference of 0.05 dex ($\sigma$ = 0.03 dex) and
for the microturbulent velocities we found 0.01 km s$^{-1}$ ($\sigma$
= 0.04 km s$^{-1}$).  We repeated the chemical analysis with the new
parameters, finding that they do not alleviate the difference between
the average [Fe I/H] and [Fe II/H] abundance ratios: we obtained [Fe I/H] $= -0.94
\pm 0.01$ dex ($\sigma$ = 0.06 dex) and [Fe I/H] $= -0.84 \pm 0.01$
dex ($\sigma$ = 0.07 dex).  Thus, the iron abundances estimated from
Fe I lines remain systematically lower than those obtained from Fe II
lines and those found in the RGB stars.

\emph{Microturbulent velocity --} We note that the (spectroscopically)
derived values of $v_{\rm turb}$ span a large range (between 1 and 2
km s$^{-1}$ for most of the targets). Also, a small trend between the
average abundances and $v_{\rm turb}$ is detected, [Fe I/H] increasing
by 0.15 dex/km s$^{-1}$ and [Fe II/H] varying by 0.08 dex/km s$^{-1}$.
The very large number of lines ($\sim$150) used to constrain $v_{\rm turb}$,
as well as the wide range of line strengths covered by the
selected transitions, ensure that no bias due to small number
statistics or small range of line strengths occurs in the
determination of $v_{\rm turb}$ (note that no specific trend between
[Fe/H] and $v_{\rm turb}$ is found among the RGB stars). Also, the
values of $v_{\rm turb}$ do not change significantly changing the
range of used reduced EWs (see Section 3.2).

We checked the impact of a different $v_{\rm turb}$ scale, adopting
the $v_{\rm turb}$--$\log g$ relation provided by \citet{kirby09}.
Because our targets have very similar gravities, they have ultimately
the same value of $v_{\rm turb}$ ($\sim$ 1.7 km s$^{-1}$), and the
situation worsens: in several stars the dispersion around the mean
abundance significantly increases (up to $\sim$0.3 dex, in comparison
with $\sigma$ = 0.15 dex found with the spectroscopic estimate of
$v_{\rm turb}$).  This is a consequence of the trends found between
abundances and line strengths introduced by not optimized $v_{\rm
turb}$.  The new average abundances of the entire sample are
[Fe I/H] $= -1.03 \pm 0.04$ dex ($\sigma$ = 0.18 dex) and [Fe II/H] $=
-0.89 \pm 0.02$ dex ($\sigma$ = 0.12 dex). Hence, with a different
assumption about $v_{\rm turb}$ not only the star-to-star dispersion
increases by a factor of 2 for both the abundance ratios, but, also,
the systematic difference between [Fe I/H] and [Fe II/H] remains in
place.

\emph{Model atmospheres --} The plane-parallel geometry
is adopted both in the ATLAS9 model atmospheres and 
in the line-formation calculation performed by GALA.
As pointed out by \citet{heiter06}, that investigated the impact 
of the geometry on the abundance analysis of giant stars, 
the geometry has a small effect on line formation. 
In order to quantify these effects, we reanalyzed the target stars by using 
the last version of the MARCS model atmospheres \citep{gustafsson08},
which adopt spherical geometry. 
The average abundance differences between the analysis performed with 
MARCS and that performed with ATLAS9 are of --0.005 dex ($\sigma$ = 0.01 dex)
and +0.02 dex ($\sigma$ = 0.04 dex) for Fe~I and Fe~II, respectively.
Hence, the use of MARCS model atmospheres does not change our finding
about Fe~I and Fe~II abundances (both in AGB and in RGB stars). 
Note that \citet{heiter06} conclude that abundances derived 
with spherical models and plane-parallel transfer are in excellent agreement 
with those obtained with a fully spherical treatment.

\section{Discussion}
\label{discussion}

\subsection{A possible signature of NLTE effects?}

For the 24 AGB stars studied in 47 Tuc, the iron abundance obtained
from single ionized lines well matches that measured in RGB stars
(from both Fe I and Fe II lines).  Instead, systematically lower iron
abundances are found for the AGB sample from the analysis of Fe I.
All the checks discussed in Section \ref{check} confirm that such a
discrepancy is not due to some bias in the analysis or to the adopted
atmospheric parameters, and there are no ways to reconcile the
abundances from Fe lines with those observed in the RGB
stars.

The only chemical analyses performed so far on AGB stars in 47 Tuc
have been presented by \citet{wylie06} and \citet{worley09}. In both
cases all the parameters have been constrained spectroscopically (in
particular, $\log g$ is obtained by forcing [Fe I/H] and [Fe II/H] to
be equal within the uncertainties).  \citet{wylie06} analysed 5 AGB
stars (brighter than those discussed in this work), finding [Fe I/H]
$= -0.60 \pm 0.06$ dex and [Fe II/H] =$ -0.64 \pm 0.10$ dex.  The same
methodology to derive the parameters has been used by \citet{worley09}
to analyse a bright AGB star, finding [Fe I/H]$ = -0.72 \pm 0.16$ dex
and [Fe II/H]$ = -0.74 \pm 0.08$ dex.  Unfortunately, the spectroscopic
determination of the gravity does not allow to understand whether also
for these AGB stars a real discrepancy of [Fe I/H] and [Fe II/H] does
exist.

A natural explanation for the negative values of [Fe I/H]$-$[Fe II/H]
measured for our AGB sample would be that these stars suffer for
departures from the LTE condition, which mainly affects the less
abundant species (in this case Fe I), while leaving virtually
unaltered the dominant species \citep[i.e. Fe II;][]{mashonkina11}.
In late-type stars, NLTE effects are mainly driven by overionization
mechanisms, occurring when the intensity of the radiation field
overcomes the Planck function \citep[see][for a complete review of
  these effects]{asplund05}.  These effects are predicted to increase
for decreasing metallicity and for decreasing atmospheric densities
(i.e., lower surface gravities at a given $T_{\rm eff}$), as pointed
out by a vast literature
\citep[see e.g.][]{thevenin99,asplund05,mashonkina11,lind12,bergemann14}.
At the metallicity of 47 Tuc, significant deviations are expected only
for stars approaching the RGB-Tip. \citet{bergemann12} and
\citet{lind12} computed a grid of NLTE corrections for a sample of Fe
I and Fe II lines in late-type stars over a large range of metallicity.
Assuming the atmospheric parameters of the 11 RGB stars in our sample
and the measured EWs of the iron lines in common with their grid (25
Fe I and 9 Fe II lines), the predicted NLTE corrections are 
[Fe/H]$_{\rm NLTE}-$[Fe/H]$_{\rm LTE}\simeq +0.04$ dex.  This is
consistent with no significant differences between [Fe I/H] and [Fe
  II/H] found in our analysis (Section \ref{check}) and in previous
studies \citep[see e.g.][]{carretta04,koch08,carretta09}.  Instead, a
larger difference ([Fe I/H]$-$[Fe II/H]$ = -0.08$ dex) has been found
for the brightest RGB stars in 47 Tuc \citep{koch08}, as expected.
\footnote{For sake of comparison, \citet{mucciarelli13c} analysed RGB
  stars close to the RGB Tip in the metal-poor GC NGC 5694 ([Fe/H]
  $\sim -2.0$ dex), finding an average difference between [Fe I/H] and
  [Fe II/H] of $-0.14$ dex, consistent with the expected
  overionization effects.}

However, if we use the same grid to estimate the NLTE corrections for
our sample of AGB stars, we find [Fe/H]$_{\rm NLTE}-$[Fe/H]$_{\rm LTE}= +0.06$ dex.
This value is consistent with the NLTE corrections
predicted for RGB stars and smaller than the difference we observe between
Fe I and Fe II in our sample of AGB stars.
Interestingly, a situation similar to that encountered in
the present work has been met by \citet{ivans01} in the spectroscopic
analysis of giant stars in the GC M~5.  Their sample includes 6 AGB and
19 RGB stars, ranging from the luminosity level of the AGB clump, up
to the RGB-Tip. Also in their analysis, [Fe I/H] in AGB stars is
systematically lower (by about 0.15 dex) than [Fe II/H], while no
differences are found for the RGB stars.  The authors performed
different kinds of analysis, finding that the only way to reconcile
the iron abundance in AGB stars with the values obtained in RGB
stars is to adopt the photometric gravities and rely on the Fe II
lines only, which are essentially insensitive to LTE departures.  Our
findings, coupled with the results of \citet{ivans01} in M~5, suggest
that the NLTE effects could depend on the evolutionary stage (being
more evident in AGB stars with respect to RGB stars), also at
metallicities where these effects should be negligible (like in the
case of 47 Tuc that is more metal-rich than M~5).
This result is somewhat surprising, because a
dependence of NLTE effects on the evolutionary stage is not
expected by the theoretical models.

\begin{figure}[]
\includegraphics[trim=0cm 0cm 0cm 0cm,clip=true,scale=.60,angle=270]{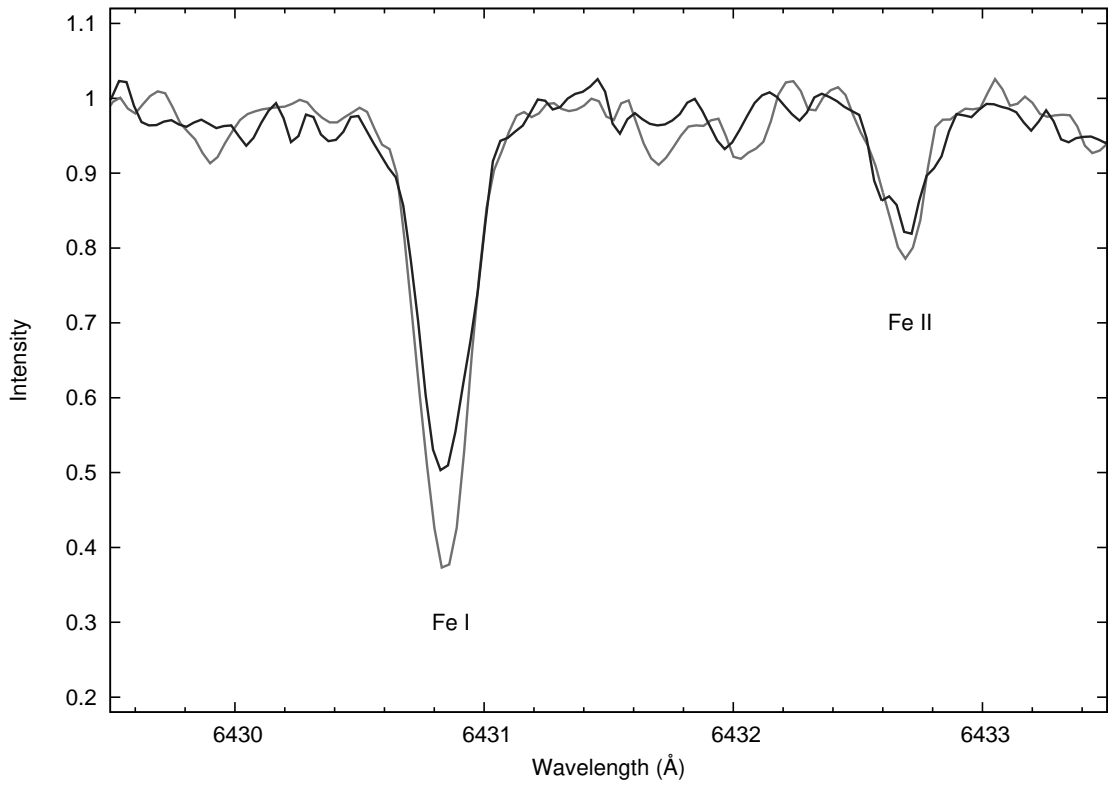}
\caption{Comparison between the normalized spectra of target \#100171
  (grey line) and \#100174 (black line). The position of a Fe I line
  and a Fe II line is marked.}
\label{spec}
\end{figure}

In our sample, we identify 4 AGB stars (namely \#100169, \#100171, \#200021 and \#200023)
where the absolute difference between [Fe I/H] and [Fe II/H] is quite small
(less than $0.05$ dex; see Fig. \ref{fediff}). According to the different behaviour observed
between the AGB and the RGB samples, one could suspect that these
objects are RGB stars. However, we checked their position on the CMD
also by using an independent photometry \citep{sarajedini07},
confirming that these 4 targets are indeed genuine AGB stars.
Fig. \ref{spec} compares two iron lines of the spectra of targets \#100171 and \#100174,
which are located in the same position of the CMD (thus being
characterized by the same atmospheric parameters), 
but have [Fe I/H]$-$[Fe II/H]$ = 0$ and $-0.16$ dex, respectively. Clearly, the Fe
II lines of the two stars have very similar depths, suggesting the
same iron abundance, while the Fe I line of \#100174 (black spectrum) is significantly
shallower than that in the other star.  This likely suggests that NLTE
effects among the AGB stars have different magnitudes, the
overionization being more or less pronounced depending on the star.

The origin of this behaviour, as well as the unexpected occurrence of
NLTE effects in AGB stars of these metallicity and atmospheric
parameters, are not easy to interpret and their detailed investigation
is beyond the scope of this paper.  Suitable theoretical models of the
line formation under NLTE conditions in AGB stars should be computed
in order to explain the observed difference in the Fe I and Fe II
abundances.  We cannot exclude that some
inadequacies of the 1-dimensional model atmospheres can play a role in
the derived results. Up to now, 3-dimensional hydrodynamics
simulations of the convective effects in AGB stars have been performed
only for a typical AGB star during the thermal pulses phase and with
very low $T_{\rm eff}$, $\sim$2800 K \citep{freytag}. Similar
sophisticated models for earlier and warmer phases of the AGB are
urged.

\subsection{Impact on traditional chemical analyses}
\label{tradit}

This result has a significant impact on the approach traditionally
used for the chemical analysis of AGB stars.  In particular, two main
aspects deserve specific care:

(1) the Fe I lines should not be used to determine the iron abundance
of AGB stars. In fact, when photometric parameters are adopted,
the [Fe I/H] abundance ratio can be systematically lower than that
obtained from Fe II lines.  Indeed, the most reliable route to derive
the iron abundance in AGB stars is to use the Fe II lines, that are
essentially unaffected by NLTE effects and provide the same abundances
for both RGB and AGB objects. This strategy requires high-resolution,
high-quality spectra, because of the low number of Fe II lines
available in the optical range (smaller by a factor of $\sim 10$ with
respect to the number of Fe I lines).  This is especially true 
at low metallicity since the lines are shallower and the NLTE effects 
are expected to be stronger;

(2) another point of caution is that AGB stars must be analysed by
adopting the photometric gravities (derived from a theoretical
isochrone or through the Stefan-Boltzmann equation) and not by using
the spectroscopic method of the ionization balance (at variance with
the case of RGB stars, where this approach is still valid).  This
method, which is widely adopted in the chemical analysis of optical
stellar spectra, constrains $\log g$ by imposing 
the same abundance for given species as obtained from
lines of two different ionization stages.  Variations of $\log g$ lead
to variations in the abundances measured from ionized lines (which are
very sensitive to the electronic pressure), while the neutral lines
are quite insensitive to these variations.  Because of the
systematically lower Fe I abundance, this procedure leads to
improbably low surface gravities in AGB stars (as demonstrated in
Section~\ref{check}).

If the Fe I line are used as main diagnostic of the iron abundance, a
{\sl blind} analysis, where AGB and RGB stars are not analysed
separately, can lead to a spurious detection of large iron spreads in
GCs.  In light of these considerations, the use of Fe II lines is
recommended to determine the metallicity of AGB objects, regardless of
their metallicity and luminosity.  On the other hand, the systematic
difference between Fe I abundances in AGB and RGB stars could, in
principle, be used to recognize AGB stars when reliable photometric
selections cannot be performed.

\subsection{Searching for evolved BSSs among AGB stars: a new diagnostic?}
\label{ebss}
The discovery of such an unexpected NLTE effect in AGB stars might
help to identify possible e-BSSs among AGB stars.  In fact, BSSs spend
their RGB phase in a region of the CMD which is superimposed to that
of the cluster AGB \citep[e.g.,][]{beccari06,dalessandro09}.  Thus, e-BSSs are
indistinguishable from genuine AGB stars in terms of colors and
magnitudes, but they have larger masses at comparable radii.  Hence,
their surface gravity is also expected to be larger (by about 0.2-0.3
dex) than that of ``canonical'' AGB stars of similar temperature and
luminosity.  Unfortunately, the spectroscopic measurement of $\log g$
cannot be used to distinguish between genuine AGB stars and e-BSSs,
since the ionization balance method cannot be applied in the presence
of the NLTE effects affecting the AGB (see Sect. \ref{tradit}).
However, because of the seemingly dependence of the NLTE effects on
the evolutionary phase, the measurement of different 
values of [Fe I/H] and [Fe II/H] should allow to recognize genuine AGB
stars from e-BSSs evolving along their RGB within a sample of objects
observed in the AGB clump of a GC.

Of course, this expectation holds only if the lack of NLTE
effects found for low-mass ($\sim 0.8~M_\odot$) RGB stars also holds
for larger masses ($\sim 1.2~M_\odot$), typical of e-BSSs in
GCs.  In order to check this hypothesis, we retrieved from the ESO
archive FLAMES-UVES spectra for 7 giant stars in the open cluster
Berkeley 32 (ID Program: 074.D-0571). This cluster has an age of $\sim$6-7 Gyr
\citep{dorazi06}, corresponding to a turnoff mass of $\sim
1~M_{\odot}$, comparable with the typical masses of the BSSs observed
in Galactic GCs
\citep[e.g.,][]{shara97,demarco05,fe06_COdep,lan07_1904,fiorentino14}.
We analyzed the spectra following the same procedure used for the AGB
stars in 47 Tuc.  The derived abundances are [Fe I/H]$ = -0.34 \pm
0.01$ dex ($\sigma$ = 0.04 dex) and [Fe II/H]$ = -0.38 \pm 0.02$ dex
($\sigma$ = 0.06 dex), in nice agreement with the results of
\citet{sestito06} based on the same dataset. The small difference
between the abundances from Fe I and Fe II lines confirms the evidence
arising from the RGB sample of 47 Tuc: in metal-rich RGB stars no sign
of overionization is found, also for stellar masses larger than those
typical of Galactic GC stars.

Within this framework, we checked whether some e-BSSs could be hidden
in the analysed sample of putative AGB stars, especially among the
four objects with negligible difference between Fe I and Fe II
abundances. To this end we derived new atmospheric parameters for all
targets projecting their position in the CMD on a grid of evolutionary
BaSTI tracks crossing the mean locus of the studied stars: these are
the RGB tracks for stellar masses of 1.2, 1.3 and 1.4~$M_\odot$
\citep[see][]{beccari06}. On average, the new values of $T_{\rm eff}$
are slightly larger than those obtained in Section \ref{chem}, while
gravities are systematically larger, by up to $\sim$0.3 dex.  For
increasing mass of the adopted track, the general behaviour is that
[Fe I/H] slightly increases (mainly because of a small increase of
$T_{\rm eff}$), while [Fe II/H] decreases (because of the combined
growth of both $T_{\rm eff}$ and $\log g$).  However, none of the measured 
AGB stars show [Fe I/H]$\sim$[Fe II/H]$\sim$[Fe/H]$_{\rm RGB}$ with the new
sets of parameters. This suggests that no e-BSSs are hidden within our
observed AGB sample.  In particular, we found that for the four AGB
stars with no evidence of overionization, the [Fe II/H] abundance
ratios derived with the new parameters are larger than those obtained
from [Fe I/H] (due to the increase in $\log g$) and for RGB stars.
This indicates that the atmospheric parameters estimated from the AGB
portion of the isochrone are the most appropriate and these four
objects are indeed genuine AGB stars.  As an additional check, we
measured the iron abundances assuming atmospheric parameters from
theoretical tracks of massive (1-2~$M_\odot$) AGB stars. In this case,
the situation improves.  In particular, by using an AGB track of 1.2~$M_\odot$, 
because of the combined effect of larger $T_{\rm eff}$
($\sim$+100 K) and $\log g$ ($\sim$+0.2), we find 
average values of [Fe I/H]$ = -0.85$ dex and [Fe II/H]$ = -0.81$ dex.
However, this scenario is unlikely because the probability
to detect an e-BSS during its AGB phase is a factor of $\sim 10$ lower
than the probability to observe it during the RGB phase (consistently
with the time duration of these evolutionary phases).

As a general rule, however, we stress that, the dependence of NLTE
effects on the evolutionary stage (if confirmed) can be used as a
diagnostic of the real nature of the observed AGB stars and to
identify e-BSSs hidden in a putatively genuine AGB sample.

\section{Summary}
\label{summary}

We have measured the iron abundance of 24 AGB stars members of the GC
47 Tuc, by using high-resolution FEROS spectra.  By adopting photometric
estimates of $T_{\rm eff}$ and $\log g$, we derived average iron
abundances [Fe I/H]$ = -0.94 \pm 0.01$ dex ($\sigma$ = 0.08 dex) and
[Fe II/H]$ = -0.83 \pm 0.01$ dex ($\sigma$ = 0.05 dex).  Thus, while
the abundance estimated from ionized lines ($-0.83$ dex) well matches
the one obtained for RGB stars, the values measured from neutral lines
appear to be systematically lower.
We carefully checked all the steps of our
chemical analysis procedure and the adopted atmospheric parameters,
finding no ways to alleviate this discrepancy.

Such a difference is compatible with the occurrence of NLTE effects
driven by iron overionization, confirming the previous claim by
\citet{ivans01} for a sample of AGB stars in M~5.  Our findings suggest that
$(i)$ the departures from the LTE approximation can be more
significant than previously thought, even at relatively high
metallicities ([Fe/H]$\sim -0.83$) and for stars much fainter
than the RGB-Tip, $(ii)$ iron overionization can be more or less
pronounced depending on the star (in fact, four stars in our sample
turn out to be unaffected), and $(iii)$ these effects depend on the
evolutionary stage (they are not observed among RGB stars).  We
discussed the impacts of this effect on the traditional chemical
analysis of AGB stars: if Fe I lines are used and/or surface gravities
are derived from the ionization balance, artificial under-estimates
and/or spreads of the iron abundances can be obtained.  If the
dependence of these NLTE effects on the evolutionary stage is
confirmed, the systematic difference between [Fe I/H] and [Fe II/H]
abundance ratios can, in principle, be used to identify e-BSSs within
a sample of genuine AGB stars (these two populations sharing the same
locus in the CMD).

From the theoretical point of view, new and accurate models are needed
to account for these findings, in particular to explain the dependence
on the evolutionary stage.  Observationally, further analyses of
high-resolution spectra of AGB stars are crucial and urged to firmly
establish the occurrence of these effects and to investigate their
behaviour as a function of other parameters, like the cluster
metallicity, the stellar mass and the stellar luminosity.

\acknowledgements  
This research is part of the project Cosmic-Lab (see http://www.cosmic-lab.eu)
funded by the European Research Council (under contract ERC-2010-AdG-267675).
The authors thank Piercarlo Bonifacio, Thibault Merle, Elena Pancino and Monique Spite for the 
useful discussions.
We warmly thank the anonymous referee for his/her suggestions in improving the paper.


\begin{deluxetable}{rrrccr}
\tablecolumns{6}
\tiny
\tablewidth{0pt}
\tablecaption{Photometric properties and radial velocities of the AGB sample}
\tablehead{\colhead{ID} & \colhead{RA} & \colhead{Dec} & \colhead{$m_{\rm F606W}$}
& \colhead{$m_{\rm F814W}$} & \colhead{RV} \\
& (J2000) & (J2000) & \colhead{} & \colhead{} & \colhead{(km s$^{-1}$)}  }
\startdata
 & & & & & \\
 100094 & 6.1013041 & --72.0745785 & 12.50 & 11.61 & --28.20 $\pm$ 0.02 \\
 100103 & 6.0415972 & --72.0787949 & 12.58 & 11.75 & --19.03 $\pm$ 0.03 \\
 100110 & 6.0212020 & --72.0791089 & 12.61 & 11.76 & --24.71 $\pm$ 0.02 \\
 100115 & 6.0253911 & --72.0763508 & 12.66 & 11.81 &   +9.42 $\pm$ 0.03 \\
 100118 & 6.0021239 & --72.0797989 & 12.66 & 11.83 & --22.55 $\pm$ 0.03 \\
 100119 & 5.9937010 & --72.1048573 & 12.66 & 11.82 &  --0.83 $\pm$ 0.02 \\
 100120 & 6.0686535 & --72.0977710 & 12.67 & 11.83 & --36.76 $\pm$ 0.02 \\
 100125 & 6.0229324 & --72.0843986 & 12.68 & 11.85 & --28.72 $\pm$ 0.03 \\
 100133 & 6.0062382 & --72.0914060 & 12.74 & 11.92 & --12.33 $\pm$ 0.02 \\
 100136 & 6.0116940 & --72.0848407 & 12.75 & 11.94 & --13.21 $\pm$ 0.02 \\
 100141 & 6.0474465 & --72.0917594 & 12.77 & 11.93 & --13.46 $\pm$ 0.02 \\
 100142 & 6.0474090 & --72.1034811 & 12.77 & 11.93 & --20.41 $\pm$ 0.03 \\
 100148 & 6.0283392 & --72.0802692 & 12.80 & 11.99 & --26.31 $\pm$ 0.03 \\
 100151 & 5.9726093 & --72.1059031 & 12.81 & 12.00 &  --8.50 $\pm$ 0.03 \\
 100152 & 6.0354605 & --72.0975184 & 12.81 & 11.97 &  --2.45 $\pm$ 0.02 \\
 100154 & 6.0171898 & --72.0853225 & 12.82 & 12.00 &  --4.70 $\pm$ 0.03 \\
 100161 & 6.0075116 & --72.0971679 & 12.84 & 12.00 & --22.37 $\pm$ 0.02 \\
 100162 & 6.0212298 & --72.0768974 & 12.85 & 12.01 & --41.45 $\pm$ 0.03 \\
 100167 & 6.0406724 & --72.0919126 & 12.87 & 12.03 & --13.98 $\pm$ 0.04 \\
 100169 & 6.0416117 & --72.1082069 & 12.87 & 12.04 & --21.38 $\pm$ 0.02 \\
 100171 & 6.0479798 & --72.0906284 & 12.88 & 12.05 & --12.69 $\pm$ 0.02 \\
 100174 & 6.0424039 & --72.0857888 & 12.90 & 12.07 & --13.81 $\pm$ 0.03 \\
 200021 & 6.1149444 & --72.0892243 & 12.74 & 11.89 & --21.07 $\pm$ 0.02 \\
 200023 & 6.1184182 & --72.0838295 & 12.85 & 12.02 & --22.19 $\pm$ 0.02 \\
\enddata
\tablecomments{Identification number, coordinates, $m_{\rm F606W}$ and $m_{\rm F814W}$
magnitudes \citep{beccari06}, and radial velocities for the 24
AGB stars analyzed.}
\label{tab1}
\end{deluxetable}

\begin{deluxetable}{cccccccc}
\tablecolumns{8}
\tiny
\tablewidth{0pt}
\tablecaption{Atmospheric parameters and iron abundances of the AGB sample}
\tablehead{\colhead{ID} & \colhead{$T_{\rm eff}^{\rm phot}$} & \colhead{$\log g^{\rm phot}$}
&  \colhead{$v_{\rm turb}$}
& \colhead{[Fe I/H]} & \colhead{n(Fe I)} & \colhead{[Fe II/H]} & \colhead{n(Fe II)} \\
& \colhead{(K)} & \colhead{(dex)} &  \colhead{(km s$^{-1}$)} & \colhead{(dex)} & & \colhead{(dex)} & }
\startdata
 & & & & & & &  \\
 100094  &  4425  &   1.55  &    2.00  &  --0.91$\pm$0.04  &  134  &  --0.79$\pm$0.08  &  13  \\
 100103  &  4450  &   1.60  &    1.15  &  --1.01$\pm$0.05  &  170  &  --0.76$\pm$0.10  &  14  \\
 100110  &  4475  &   1.60  &    1.10  &  --0.98$\pm$0.04  &  165  &  --0.85$\pm$0.07  &  12  \\
 100115  &  4500  &   1.65  &    0.55  &  --0.99$\pm$0.05  &  171  &  --0.84$\pm$0.09  &  13  \\
 100118  &  4500  &   1.65  &    1.30  &  --0.98$\pm$0.05  &  161  &  --0.85$\pm$0.08  &  15  \\
 100119  &  4500  &   1.65  &    1.80  &  --0.85$\pm$0.04  &  138  &  --0.71$\pm$0.07  &  15  \\
 100120  &  4500  &   1.65  &    1.70  &  --0.97$\pm$0.05  &  138  &  --0.74$\pm$0.07  &  13  \\
 100125  &  4500  &   1.65  &    0.95  &  --1.08$\pm$0.04  &  171  &  --0.94$\pm$0.08  &  14  \\
 100133  &  4550  &   1.70  &    1.50  &  --0.98$\pm$0.04  &  147  &  --0.83$\pm$0.08  &  14  \\
 100136  &  4550  &   1.70  &    1.30  &  --0.93$\pm$0.04  &  166  &  --0.84$\pm$0.07  &  15  \\
 100141  &  4550  &   1.70  &    1.65  &  --0.93$\pm$0.04  &  155  &  --0.81$\pm$0.07  &  11  \\
 100142  &  4550  &   1.70  &    1.60  &  --0.90$\pm$0.05  &  140  &  --0.80$\pm$0.07  &  12  \\
 100148  &  4575  &   1.75  &    0.95  &  --1.05$\pm$0.04  &  173  &  --0.89$\pm$0.08  &  13  \\
 100151  &  4575  &   1.75  &    1.85  &  --0.90$\pm$0.05  &  141  &  --0.82$\pm$0.07  &  12  \\
 100152  &  4575  &   1.75  &    1.80  &  --0.91$\pm$0.04  &  156  &  --0.78$\pm$0.07  &  14  \\
 100154  &  4575  &   1.75  &    1.20  &  --1.00$\pm$0.04  &  159  &  --0.89$\pm$0.07  &  15  \\
 100161  &  4575  &   1.75  &    1.40  &  --0.90$\pm$0.05  &  158  &  --0.84$\pm$0.08  &  13  \\
 100162  &  4600  &   1.75  &    0.75  &  --1.02$\pm$0.05  &  174  &  --0.91$\pm$0.07  &  13  \\
 100167  &  4600  &   1.80  &    1.20  &  --0.95$\pm$0.05  &  158  &  --0.77$\pm$0.09  &  13  \\
 100169  &  4600  &   1.80  &    1.80  &  --0.79$\pm$0.04  &  154  &  --0.79$\pm$0.07  &  14  \\
 100171  &  4600  &   1.80  &    1.60  &  --0.83$\pm$0.04  &  155  &  --0.83$\pm$0.07  &  11  \\
 100174  &  4600  &   1.80  &    1.10  &  --1.02$\pm$0.06  &  144  &  --0.86$\pm$0.08  &  13  \\
 200021  &  4550  &   1.70  &    1.90  &  --0.83$\pm$0.04  &  145  &  --0.84$\pm$0.07  &  14  \\
 200023  &  4575  &   1.75  &    1.80  &  --0.81$\pm$0.04  &  147  &  --0.79$\pm$0.07  &  15  \\
\hline
 &  &  &  & $\langle$[Fe I/H]$\rangle$&  & $\langle$[Fe II/H]$\rangle$ &   \\
 &  &  &  & --0.94$\pm$0.01  &  & --0.83$\pm$0.01  &  \\
\enddata
\tablecomments{ Identification number, photometric temperature and
  gravities, microturbulent velocities, [Fe/H] abundance ratios with
  total uncertainty and number of used lines, as measured from 
  neutral and single ionized lines.  For all the stars a global
  metallicity of [M/H]$ = -1.0$ dex has been assumed for the model
  atmosphere. The adopted solar value is 7.50 \citep{grevesse98}.}
\label{tab2}
\end{deluxetable}

\begin{deluxetable}{cccccccc}
\tablecolumns{8}
\tiny
\tablewidth{0pt}
\tablecaption{Atmospheric parameters and iron abundances of the RGB sample}
\tablehead{\colhead{ID} & \colhead{$T_{\rm eff}^{\rm phot}$} & \colhead{$\log g^{\rm phot}$}
&  \colhead{$v_{\rm turb}$}
& \colhead{[Fe I/H]} & \colhead{n(Fe I)} & \colhead{[Fe II/H]} & \colhead{n(Fe II)} \\
& \colhead{(K)} & \colhead{(dex)} &  \colhead{(km s$^{-1}$)} & \colhead{(dex)} & & \colhead{(dex)} & }
\startdata
 & & & & & & &  \\
 5270  &  4035  &  1.10  &  1.50  &  --0.85$\pm$0.05  &  140  &  --0.81$\pm$0.13  &  12 \\
12272  &  4130  &  1.25  &  1.50  &  --0.87$\pm$0.05  &  147  &  --0.86$\pm$0.11  &  13 \\
13795  &  4170  &  1.35  &  1.60  &  --0.81$\pm$0.05  &  141  &  --0.79$\pm$0.11  &  14 \\
14583  &  4305  &  1.60  &  1.50  &  --0.81$\pm$0.05  &  150  &  --0.81$\pm$0.10  &  13 \\
17657  &  4005  &  1.05  &  1.50  &  --0.86$\pm$0.04  &  133  &  --0.90$\pm$0.10  &  12 \\
18623  &  4250  &  1.50  &  1.50  &  --0.84$\pm$0.05  &  144  &  --0.86$\pm$0.10  &  12 \\
20002  &  4200  &  1.40  &  1.50  &  --0.86$\pm$0.05  &  147  &  --0.84$\pm$0.10  &  12 \\
23821  &  4250  &  1.50  &  1.20  &  --0.84$\pm$0.04  &  147  &  --0.84$\pm$0.09  &  14 \\
34847  &  4095  &  1.20  &  1.40  &  --0.82$\pm$0.05  &  141  &  --0.82$\pm$0.12  &  13 \\
36828  &  4215  &  1.40  &  1.40  &  --0.78$\pm$0.05  &  142  &  --0.84$\pm$0.11  &  11 \\
41654  &  4130  &  1.25  &  1.50  &  --0.82$\pm$0.05  &  142  &  --0.85$\pm$0.11  &  13 \\
\hline
 &  &  &  & $\langle$[Fe I/H]$\rangle$&  & $\langle$[Fe II/H]$\rangle$ &   \\
 &  &  &  & --0.83$\pm$0.01  &  & --0.84$\pm$0.01  &  \\
\enddata
\tablecomments{Columns are as in Table 2. For all the stars a global
metallicity of [M/H]$ = -1.0$ dex has been assumed for the model
atmosphere. The adopted solar value is 7.50 \citep{grevesse98}.}
\label{tab3}
\end{deluxetable}

\end{document}